\documentclass[aoas,MSNbibl,nameyear,dvips]{arximspdf}
\usepackage{dcolumn}
\usepackage{graphicx}
\doi{10.1214/13-AOAS705} 
\volume{8}
\issue{1}
\pubyear{2014}
\firstpage{148}
\lastpage{175}

\makeatletter
\newcolumntype{d}[1]{D{.}{.}{#1}}
\newcommand{\eqref}[1]{(\ref{#1})}

\def\n{\nu}

\def\I{\mathbf{I}}
\def\A{\mathbf{A}}
\def\Yg{\mathbf{Y}_g}
\def\R{\mathbf{R}}
\def\Ug{\mathbf{U}_g}
\def\xiR{\bolds{\xi}_R}
\def\xiRT{\bolds{\xi}'_R}
\def\cb{c_{\beta}}
\def\cmu{c_{\mu}}
\def\Hs{\mathbf{H}_n}

\def\Xbar{\bar{X}}
\def\mun{o_{h1}}
\def\md{o_{h2}}
\def\mt{o_{h3}}
\def\mq{o_{h4}}

\newcommand{\Bern}{\operatorname{Bern}}

\makeatother

\begin{document}
\begin{frontmatter}

\title{A hierarchical Bayesian model for
inference of copy number variants and their association to gene expression}
\runtitle{Bayesian modeling for gene-CNV association and inference}

\begin{aug}
\author[a]{\fnms{Alberto} \snm{Cassese}\thanksref{m1}\ead[label=e1]{Alberto.Cassese@rice.edu}},
\author[b]{\fnms{Michele} \snm{Guindani}\thanksref{m2}\ead[label=e2]{mguindani@mdanderson.org}},
\author[c]{\fnms{Mahlet G.} \snm{Tadesse}\thanksref{m3}\ead[label=e3]{mgt26@georgetown.edu}},
\author[d]{\fnms{Francesco} \snm{Falciani}\thanksref{m4}\ead[label=e4]{f.falciani@liverpool.ac.uk}}
\and
\author[a]{\fnms{Marina} \snm{Vannucci}\corref{}\thanksref{m1}\ead[label=e5]{marina@rice.edu}}
\runauthor{A. Cassese et al.}
\affiliation{Rice University\thanksmark{m1},
MD Anderson Cancer Center\thanksmark{m2},
Georgetown University\thanksmark{m3}\break
and University of Liverpool\thanksmark{m4}}
\address[a]{A. Cassese\\
M. Vannucci\\
Department of Statistics\\
Rice University\\
Houston, Texas 77005\\
USA\\
\printead{e1}\\
\phantom{E-mail:\ }\printead*{e5}}

\address[b]{M. Guindani\\
Department of Biostatistics\\
MD Anderson Cancer Center\\
Houston, Texas 77030\\
USA\\
\printead{e2}\hspace*{12pt}}

\address[c]{M. G. Tadesse\\
Department of Mathematics and Statistics\\
Georgetown University\\
Washington, DC 20057\\
USA\\
\printead{e3}}

\address[d]{F. Falciani\\
Center of Computational Biology\\
\quad and Modelling (CCMB)\\
Institute of Integrative Biology\\
University of Liverpool\\
Liverpool\\
United Kingdom\\
\printead{e4}}
\end{aug}

\received{\smonth{4} \syear{2013}}
\revised{\smonth{11} \syear{2013}}

%
\begin{abstract}
A number of statistical models have been successfully developed for the
analysis of high-throughput data from a single source, but few methods
are available for integrating data from different sources. Here we
focus on integrating gene expression levels with comparative genomic
hybridization (CGH) array measurements collected on the same subjects.
We specify a measurement error model that relates the gene expression
levels to latent copy number states which, in turn, are related to the
observed surrogate CGH measurements via a hidden Markov model. We
employ selection priors that exploit the dependencies across adjacent
copy number states and investigate MCMC stochastic search techniques
for posterior inference. Our approach results in a unified modeling
framework for simultaneously inferring copy number variants (CNV) and
identifying their significant associations with mRNA transcripts
abundance. We show performance on simulated data and illustrate an
application to data from a genomic study on human cancer cell lines.
\end{abstract}

%

\end{frontmatter}

\section{Introduction}\label{sec1}
Our understanding of cancer biology and the mechanisms underlying
cancer cell growth has progressed tremendously over the past decade.
Cancer is the consequence of a dynamic interplay at different molecular
levels (DNA, mRNA and protein). Elucidating the association between two
or more of these levels would enable the identification of biological
relationships that could lead to improvements in cancer diagnosis and
treatment. Consequently, studies that integrate different types of
high-throughput data are of great interest. This paper is concerned
with the integration of gene expression and copy number variant data.

Gene expression levels correspond to the relative abundance of mRNA
transcripts. These expression levels can be altered by chromosomal
aberrations, such as copy number variants (CNV). CNVs are variations in
the copy number of DNA segments due to cytogenetic events, in which the
DNA replication process is disrupted and the DNA segment is either
replicated (once or several times) or deleted in newly generated cells,
leading to local chromosomal amplifications/deletions [\citet
{Sebat:2004}]. Several experimental techniques are available for CNV
detection. The most widely used high-throughput technologies include
comparative genomic hybridization (CGH) arrays and single nucleotide
polymorphism (SNP) arrays. In this paper, we focus on the former, which
generates data as reads on thousands or millions of genomic
hybridization targets (probes) spotted on a glass surface. Regions of
relative gains or losses are identified by measuring the fluorescence
ratio of differentially labeled test and reference DNA samples
hybridized onto the array. The reference DNA is assumed to have two
copies of each chromosome. If the test sample has no copy number
aberrations, the $\log_2$ of the intensity ratio is theoretically
equal to zero.

A number of statistical methods have been developed to infer CNVs from
high-throughput array-based technologies. The most widely used rely on
hidden Markov models (HMM) [\citet{Colella:2007,Wang:2007}] and
circular binary segmentation [\citet{Venkatraman:2007}]. Other methods
based on clustering have been proposed, including a combination of
segmentation and model-based clustering [\citet{Picard:2007}] and a
Bayesian hierarchical mixture model [\citet{Cardin:2011}]. These methods
process each sample one at a time and require post-processing of the
inferred CNV calls to resolve CNV boundary variations.

In addition to CNV detection, there is often interest in identifying
variants associated with specific phenotypes or biological functions.
Most of the available methods either directly use the normalized
continuous intensity measurements without inferring copy numbers or use
the estimated copy numbers as true states, then assess the associations
using univariate tests or by performing simple linear regression models
with multiple testing correction [\citet{Stranger:2007}, \citet
{Wang:2007}]. When using the raw measurements, the aggregation of a
large number of tests with low $p$-values in close genetic proximity is
considered evidence of copy number-phenotype association. Although this
approach has the advantage of circumventing the need to infer copy
number, the high noise in the signal intensities leads to the
identification of a large number of false positives [\citet
{Breheny:2012}]. On the other hand, using the copy number calls as if
they were the true states ignores the uncertainty in the estimation
process and can introduce bias. Several methods have been proposed to
incorporate the uncertainty in copy number estimation into the
association tests [\citet{Barnes:2008}, \citet{Subirana:2011}].

In the past few years, there has been a growing interest in relating
gene expression and CNV data. Indeed, locating CNVs that affect gene
dosage is an important step in understanding biological processes
underlying various diseases. In cancer, for example, where chromosomal
aberrations are widespread due to genomic instability, discovering
amplification of oncogenes or deletion of tumor suppressors are
important steps in elucidating tumorigenesis. Earlier attempts in this
area have used Pearson correlation coefficients to evaluate
associations between raw CGH intensities and gene expression levels
mapping to the same genomic region [\citet{Bussey:2006}, \citet
{Chin:2006}]. \citet{Choi:2010} developed a double-layered mixture model
to simultaneously estimate copy numbers and evaluate the association
between each copy number probability score and the expression level of
the corresponding gene. These models perform univariate associations
between CNVs and gene expression levels on the same chromosomal region.
However, it would be expected that multiple CNVs mapping to different
genomic regions may be associated to gene regulation, a mechanism that
is part of epistasis; see \citet{Cordell:2002}.

Several multivariate statistical methods for integrating genomic data
sets have been proposed in recent years. \citet{mahlet:2009} proposed a
stochastic partitioning method to identify sets of correlated gene
expression levels and select sets of chromosomal abberations that
jointly modulate mRNA transcript abundance in the co-expressed genes.
Other authors have proposed variable selection methods in multivariate
linear regression models in the context of eQTL (expression
quantitative trait loci) analysis. Among those, \citet{Richardson:2010}
proposed mixture priors that enforce sparsity while enhancing the
detection of predictors that are associated with many responses.
Similar priors have also been studied by \citet{Boyer:2012} for eQTL analysis.

In this paper we develop an innovative statistical model that
integrates gene expression and copy number variant data. The proposed
approach provides a unified framework to simultaneously infer CNVs
across all samples and identify significant associations between copy
number states and gene expression changes. To achieve this goal, we
first specify a joint distribution of the observed gene expression and
CGH data across all samples. Using a measurement error model
formulation, we factor this joint distribution into the product of
conditionally independent submodels: an outcome model that relates the
gene expression levels to latent copy number states, and a measurement
model that relates these latent states to the observed surrogate CGH
measurements using a first order hidden Markov model (HMM).
We identify CNVs associated with gene expression changes by
incorporating a latent indicator for variable selection into the
outcome model and specifying selection priors that account for spatial
dependences between adjacent DNA segments. Our strategy for posterior
inference uses MCMC algorithms and stochastic search methods and
results in the estimation of copy number states across all samples, as
well as the selection of groups of CNVs associated with gene
expression. The model we propose allows the identification of the joint
effect of multiple CNVs on mRNA transcript abundance, rather than
assuming univariate associations. In addition, the simultaneous
evaluation of multiple gene expression levels reduces the detection of
false positive associations by borrowing information across
co-expressed genes. We show the performance of our proposed model on
simulated data. We also analyze a case study on human cancer cell
lines. Findings support the hypothesis that our approach has the
potential to discover important linkages between gene expression and cancer.

%
\begin{figure}[t]
\includegraphics{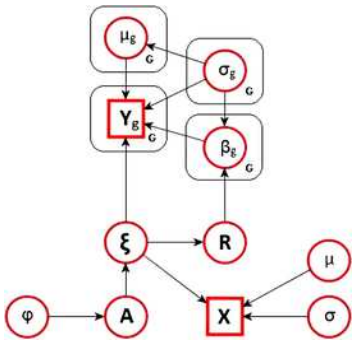}
\caption{Graphical formulation of the probabilistic model
described in Section~\protect\ref{sec:model}.}\label{fig1}
\end{figure}

The rest of the paper is organized as follows: Section~\ref{sec:model}
introduces the modeling framework and its major components and
Section~\ref{sec:pinf} describes the posterior inference and prediction.
Results on simulated data are reported in Section~\ref{sec:sim}, while
Section~\ref{sec:case} is devoted to our case study. Section~\ref
{sec:dis} contains some final remarks.

\section{Hierarchical model}
\label{sec:model}
We propose a hierarchical model that integrates gene expression levels
with copy number variant data and that accounts for the measurement
error in the observed CGH intensities via a hidden Markov model (HMM).
The model further incorporates a variable selection procedure and
utilizes prior distributions that exploit the dependencies across
adjacent DNA segments. Our modeling strategy provides a unified
approach for simultaneously inferring copy number states for all
samples and identifying associations between sets of copy number
variants and gene expression levels. The graphical formulation of the
model is illustrated in Figure~\ref{fig1} and its major components are
described below. We also summarize the hierarchical formulation of our
full model in Figure~\ref{figHM}.

\begin{figure}[t]
%
\begin{tabular}{|p{12cm}|}
\hline\noalign{\vspace*{-2pt}}\vspace*{2pt}
\textbf{Likelihood:}
\begin{eqnarray*}
f(\mathbf{Z}|\bolds{\xi}) & = & \prod_{i=1}^n
\Biggl\{\prod_{g=1}^G f(Y_{ig}|{
\bolds{\xi}}_i)\prod_{m=1}^M
f(X_{im}|\xi_{im}) \Biggr\},
\\
f(Y_{ig}|\bolds{\xi}) & = & N\bigl(\mu_g + \bolds{
\xi}_i\bolds{\beta}_g,\sigma^2_g
\bigr),
\\
f\bigl(X_{im} | (\xi_{im}=j)\bigr) & = & N\bigl(
\eta_j,\sigma^2_{j}\bigr),
\\
P(\xi_{i(m+1)}=h|\xi_{im}=j) & = & a_{hj}.
\end{eqnarray*}
\\
\hline\noalign{\vspace*{-6pt}}\vspace*{2pt}
\textbf{Model parameters:}
\begin{eqnarray*}
\beta_{gm}|r_{gm}, \sigma_g ^2 &
\sim& r_{gm} N\bigl(0,\cb^{-1} \sigma_g
^{2}\bigr) + (1-r_{gm}) \delta_{0}(
\beta_{gm}),
\\
\mu_g|\sigma^2_g &\sim& N\bigl(0,
\cmu^{-1}\sigma_g^{2}\bigr),
\\
\sigma_g^{-2} &\sim& \operatorname{Ga}(\delta/2,d/2),
\\
\eta_j &\sim& N(\delta_{j},\tau_j)
\I_{\{\mathrm{low}_{\eta_j}<\eta
_j<\mathrm{upp}_{\eta
_j}\}},
\\
\sigma_j^{-2} &\sim& \operatorname{Ga}(b_j,l_j)
\I_{\{\sigma_j^{-2}>\mathrm{upp}_{\sigma
_j}\}},
\\
\mathbf{a}_h &\sim& \operatorname{Dir}(\bolds{\phi}).
\end{eqnarray*}
\\
\hline\noalign{\vspace*{-6pt}}\vspace*{2pt}
\textbf{Variable selection parameters:}
\begin{eqnarray*}
&&\pi(r_{gm}|r_{g(m-1)},r_{g(m+1)}, \bolds{\xi},\pi_1) \\
&&\qquad= \gamma_m [\pi
_1^{r_{gm}}(1-\pi_1)^{ (1-r_{gm})}]+ \sum_{j=1}^2 \omega
^{(j)}_m I_{\{r_{gm}=r_{g(m+(-1)^j)}\}},\\
&&\pi_1 \sim\operatorname{Beta} (e,f).
\end{eqnarray*}
\\
\hline\noalign{\vspace*{-6pt}}\vspace*{2pt}
\textbf{Fixed hyperparameters:} $ \cmu, \cb, \delta, d, e, f,
\alpha,
\bolds{\delta}, \bolds{\tau}, \mathbf{b}, \mathbf{l}, \bolds
{\phi}, \mathbf
{low}_{\eta}, \mathbf{upp}_{\eta}, \mathbf{upp}_{\sigma}$\\
\hline
\end{tabular}
\caption{Hierarchical formulation of the proposed probabilistic
model.}\label{figHM}
\end{figure}

Let $Y_{ig}$ denote the expression measurement for gene $g$ $(g=1,
\ldots, G)$ and $X_{im}$ the observed CGH measurement, that is, the
normalized $\log_2$ ratio, for the $m$th CGH probe ($m=1, \ldots, M$),
in sample $i\ (i=1,\ldots,n)$. We assume the $M$ CGH probes ordered
according to their chromosomal location and refer to probes $m$ and
$m+1$ as adjacent. In our modeling approach we treat the observed CGH
intensities, $X_{im}$, as surrogates for unobserved copy number states,
which we indicate with $\xi_{im}$. Failure to account for the
measurement error, by treating the surrogates as the latent copy number
states, may lead to biased results. Here we define four copy number
states corresponding to the following:
\begin{eqnarray*}
\xi_{im}&=&1 \qquad \mbox{for copy number loss (less than two copies of the
fragment);}
\\
\xi_{im}&=&2 \qquad \mbox{for copy-neutral state (exactly two copies of the
fragment);}
\\
\xi_{im}&=&3 \qquad \mbox{for a single copy gain (exactly three copies of the
fragment);}
\\
\xi_{im}&=&4 \qquad \mbox{for multiple copy gains (more than three copies of
the fragment).}
\end{eqnarray*}

Let $\mathbf{Z}=[\mathbf{Y}, \mathbf{X}]$ denote the $(n \times
(G+M))$ matrix of
observed gene expression measurements and let $\bolds{\xi}=[\bolds
{\xi}_1,
\ldots, \bolds{\xi}_M]$ be the $(n \times M)$ matrix of latent copy number
states. We consider a nondifferential measurement error, which assumes
that, conditional on the latent state $\bolds{\xi}$, the observed
surrogate $\mathbf{X}$ contains no additional information on the response
$\mathbf{Y}$ [\citet{Richardson:1993}], that is, $f(\mathbf
{Y}|\bolds{\xi},
\mathbf{X})
= f(\mathbf{Y}|\bolds{\xi})$.
The joint distribution of $\mathbf{Z}$ can thus be decomposed into
conditionally independent submodels, that correspond to an outcome
model relating $\mathbf{Y}$ to the latent state $\bolds{\xi}$ and a
measurement model relating the surrogate $\mathbf{X}$ to $\bolds{\xi
}$, as
$f(\mathbf{Z}|\bolds{\xi}) = f(\mathbf{Y}|\bolds{\xi}) f(\mathbf
{X}|\bolds{\xi})$.
We further assume conditional independence of the gene expression
measurements, given the copy number states (i.e., $\mathbf{Y}_i
\perp
\mathbf{Y}_j | \bolds{\xi}_1,\ldots, \bolds{\xi}_M$) and conditional
independence of the CGH measurements, given their states (i.e.,
$\mathbf{X}_i \perp\mathbf{X}_j | \bolds{\xi}_1,\ldots, \bolds
{\xi
}_M$), and write
%
%
\begin{equation}
\label{jointmodel} f(\mathbf{Z}|\bolds{\xi}) = \prod_{i=1}^n
\Biggl\{\prod_{g=1}^G f(Y_{ig}|
\bolds{\xi}_i) \prod_{m=1}^M
f(X_{im}|\xi_{im}) \Biggr\}.
\end{equation}
Even though we make these assumptions, we still borrow strength across
genes via our hierarchical prior specification, as described in
Section~\ref{sec:prior}.

\subsection{Measurement error model via HMM}
\label{sec:meas_err_model}
For the outcome model in (\ref{jointmodel}) we follow \citet
{mahlet:2009} and \citet{Richardson:2010} who have suggested linear
regression models that integrate gene expression levels with genetic
data. For gene $g$ we therefore specify
a linear regression model of the type
%
%
\begin{equation}
\label{systemeqs} Y_{ig}=\mu_g+{\bolds{\xi}}_{i}{
\bolds{\beta}}_{g} + \varepsilon_{ig},\qquad i=1,\ldots,n
\end{equation}
for $g=1, \ldots, G$ and with $\mu_1,\ldots,\mu_G$ gene-specific intercepts.
We also assume $\varepsilon_{ig}\sim N(0,\sigma_g^{2})$ with $\sigma^2_g$
a gene specific variance.

We then define the measurement model in (\ref{jointmodel}) in terms of
the emission probabilities of a Hidden Markov Model (HMM).
CGH data are ``state persistent,'' meaning that copy number gains or
losses at a region are often associated to an increased probability of
gains and losses at a neighboring region. Here, we adapt the model
proposed by \citet{Guha:2008}, that uses hidden Markov models with four
copy number states. Methods that consider the number of possible states
as a random variable, such as those of \citet{Fox:2009}, \citet{Du:2010}
and \citet{Costa:2013}, may be similarly incorporated into our model.
Conditional on the latent copy number states, we assume the observed
CGH measurements independent and normally distributed, defining the
emission distributions of the HMM as
%
%
\begin{equation}
\label{modelCGH} X_{im} | (\xi_{im}=j) \stackrel{\mathrm{i.i.d.}} {\sim}
N\bigl(\eta_j,\sigma^2_{j}\bigr),
\end{equation}
with $\eta_j$ and $\sigma^2_j$ representing the expected $\log_2$ ratio
and the variance of all CGH probes in state $j$ $(j=1,\ldots,4)$. The
dependence between the states at adjacent probes is captured by a first
order Markov model, which assumes that the probability of being in a
particular copy number state at chromosomal location $m+1$ depends only
on the state at location $m$,
\[
P(\xi_{i(m+1)}|\xi_{i1}, \ldots,\xi_{im})=P(
\xi_{i(m+1)}|\xi_{im}) = a_{\xi_{im} \xi_{i(m+1)}},
\]
with $\mathbf{\A}=(a_{hj})$ forming the matrix of transition
probabilities with strictly positive elements ($h, j=1, \ldots, 4$).
This matrix has a unique stationary distribution $\pi_{A}$. The initial
probabilities of being in each of the states at $m=1$ are also assumed
to be given by $\pi_{A}$.

\subsection{Prior models for spatial dependence}
\label{sec:prior}
For each gene we wish to find a parsimonious set of CGH aberrations
that affect the gene expression levels with high confidence. This is
equivalent to inferring which elements of the vector ${\bolds{\beta
}_g}$ in
(\ref{systemeqs}) are nonzero, that is, a classical variable selection
problem. The resulting ``network'' of gene-CGH associations can be
encoded by a ($G \times M$) matrix $\R$ of binary elements.
Specifically, for gene expression $g$ and CGH probe $m$, the value
$r_{gm}=1$ indicates that the corresponding coefficient $\beta_{gm}$ is
significant, and should therefore be included in the regression model
for gene $g$. Otherwise, $r_{gm}=0$ indicates that the corresponding
regression coefficient is zero. Given $\mathbf{R}$, the regression
coefficient parameters are then stochastically independent and have the
following mixture prior distribution:
%
%
\begin{equation}
\label{prior} \pi\bigl(\beta_{gm}|r_{gm},
\sigma_g ^2\bigr)=r_{gm} N\bigl(0,
\cb^{-1} \sigma_g ^{2}\bigr) +
(1-r_{gm}) \delta_{0}(\beta_{gm}),
\end{equation}
with $\delta_{0}(\cdot)$ a point mass at zero and $\cb>0$ a
hyperparameter to be chosen (see Section~\ref{sec:sim}). The prior
model is completed with a Gamma prior on the error precision, $\sigma
_g^{-2} \sim G(\frac{\delta}{2},\frac{d}{2})$, and a Normal
distribution on the intercepts, $\mu_g|\sigma^2_g \sim N(0,\cmu
^{-1}\sigma_g^{2})$, with $\delta, d$ and $\cmu$ hyperparameters to
be chosen.

Priors of type (\ref{prior}) are known as spike-and-slab priors in the
Bayesian variable selection literature [see \citet{george:1997} for
linear regression models and \citet{brown:1998} and \citet{sha:2004} for
multivariate models] and have been employed to infer biological
networks of high dimensionality [see, e.g., \citet{jones:2005},
\citet{Richardson:2010} and \citet{Stingo:2010}]. We adopt the
formulation of \citet{Stingo:2010} which allows to select different
covariates (CNV aberrations) for different responses (genes). See also
\citet{mahlet:2009} for an approach based on partition models.

We now describe our prior choice for the elements $r_{gm}$'s of this
matrix $\R$ that encodes the association network. Since contiguous
regions of copy number changes correspond to the same DNA aberration,
they are more likely to jointly affect the expression level of a gene.
Accordingly, in our prior distribution we explicitly assume that the
probability of selection at location $m$ depends on the copy number
states and the selection of the probes at positions $\{m-1, m+1\}$.
Hence, CNVs located in regions of persistent state aberrations may be
more likely to be jointly associated with the expression levels of each
gene. We represent this dependent association structure as a
conditional mixture prior distribution
%
%
\begin{eqnarray}
\label{eq:prob_r} \pi(r_{gm}|r_{g(m-1)},r_{g(m+1)},
\bolds{\xi},\pi_1)&=&\gamma_m \bigl[\pi
_1^{r_{gm}}(1-\pi_1)^{ (1-r_{gm})}\bigr]
\nonumber
\\[-8pt]
\\[-8pt]
\nonumber
&&{}+ \sum_{j=1}^2 \omega^{(j)}_m
I_{\{r_{gm}=r_{g(m+(-1)^j)}\}},
\end{eqnarray}
where $\gamma_m\in[0,1]$ and $\sum_{j=1}^2 \omega^{(j)}_m =
(1-\gamma
_m)$. According to (\ref{eq:prob_r}), with probability $\gamma_m$, we
have that $r_{gm}\sim\Bern(\pi_1)$, independently of the neighboring
values, whereas, with probability $(1-\gamma_m)$, $r_{gm}$ coincides
with either one (or both) of the adjacent values in $\R$. We note that
equation (\ref{eq:prob_r}) reduces to the typical independence
assumption, $r_{gm}\sim \operatorname{Bern}(\pi_1)$, in the case $\gamma_m=1$.

In this paper we assume that the parameters $\gamma_m, \omega^{(1)}_m$
and $\omega^{(2)}_m$ are probe-specific, capturing information on the
physical distance between CGH probes and their unobserved copy number
states. More specifically, let $d_m$ be the distance between the
adjacent probes $\{m-1, m\}$ and let $D$ be the total length of the DNA
fragment (e.g., the length of the chromosome) under consideration. We define
%
%
\begin{equation}
\label{distance} 1-s_{(m-1)m}=1-\frac{1}{n} \sum
_{i=1} ^n \frac{e^{\{1-
{d_{m}}/{D}\}}-1}{e-1} I_{\{\xi_{im}=\xi_{i(m-1)}\}}
\end{equation}
to capture the frequency of change points at position $m$ in copy
number states across all samples. Similar quantities have been used,
for example, by \citeauthor{Wang:2007} (\citeyear
{Wang:2008,Wang:2007}) and \citet{Marioni:2006}, to
model spatial dependency in copy number detection. Here, instead, we
use them to elicit the association between each gene expression and
stretches of CNVs in the following sense. If two CGH probes are
physically close, state persistence might be more likely and the same
association pattern would be expected compared
to a situation where the two probes are located farther apart on the
genome. Accordingly, we define
%
%
\begin{eqnarray}
\label{eq:gammaomega}
\gamma_m&=&\frac{\alpha}{\alpha+s_{(m-1)m}+s_{m(m+1)}},
\nonumber
\\[-8pt]
\\[-8pt]
\nonumber
\omega^{(1)}_m&=&\frac{s_{(m-1)m}}{\alpha+s_{(m-1)m}+s_{m(m+1)}},\qquad
\omega^{(2)}_m=
\frac{s_{m(m+1)}}{\alpha+s_{(m-1)m}+s_{m(m+1)}}
\end{eqnarray}
with $\alpha$ set to a positive real value. In the applications we set
$\omega^{(1)}_m$ and $\omega^{(2)}_m$ to zero for the first and last
chromosomal locations, that is, $m=1$ and $m=M$. We note that, if
$s_{(m-1)m}=s_{m(m+1)}=0$, equation (\ref{eq:prob_r}) reduces to the
independent case, whereas larger values of either $s_{(m-1)m}$ or
$s_{m(m+1)}$ imply smaller $\gamma_m$ and, respectively, larger
$\omega
^{(1)}_m$ or $\omega^{(2)}_m$, that is, stronger spatial dependency.
The prior probability of $r_{gm} = 1$ therefore increases if
$r_{g(m-1)}$ [or $r_{g(m+1)}$] is equal to one and if there are more
samples with no change between the copy number states at locations $m$
and $m-1$ (or $m+1$).
Finally, we complete prior (\ref{eq:prob_r}) by further imposing a Beta
hyperprior, $\pi_1 \sim\operatorname{Beta} (e,f)$. Integrating $\pi
_1$ out, we obtain
%
%
\begin{eqnarray}
\label{eq:prob_r_int} %
\pi(r_{gm}|r_{g(m-1)},r_{g(m+1)},
\bolds{\xi})&=&\gamma_m \frac
{\Gamma
(e+f)\Gamma(e+r_{gm})\Gamma(f+1-r_{gm})}{\Gamma(e+f+1)\Gamma
(e)\Gamma
(f)}
\nonumber
\\[-8pt]
\\[-8pt]
\nonumber
&&{} +  \sum_{j=1}^2 \omega^{(j)}_m
I_{\{r_{gm}=r_{g(m+(-1)^j)}\}}.
\end{eqnarray}
It is immediate to show that this prior is proper since it is
nonnegative and has finite support.

As for the prior specification of the HMM of equation (\ref{modelCGH}),
we assume independent Dirichlet priors across the rows of the
transition matrix $\A$, that is, $\mathbf{a}_h = (a_{h1}, a_{h2}, a_{h3},
a_{h4}) \sim\operatorname{Dir}(\phi_1, \phi_2, \phi_3, \phi_4)$,
for $h=1,
\ldots, 4$.
For $\eta_j$ and $\sigma_j^2$ in the emission distributions (\ref
{modelCGH}) we follow \citet{Guha:2008} and assume
$\eta_j \sim N(\delta_j,\tau^2_j)I_{\{\mathrm{low}_{\eta_j}<\eta
_j<\mathrm{upp}_{\eta
_j}\}}$ and $\sigma_j^{-2} \sim\operatorname{Gamma}(b_j,l_j)I_{\{
\sigma
_j^{-2}>\mathrm{upp}_{\sigma_j}\}}$, for $j=1, \ldots, 4$. Here $\mathrm{low}_{\eta
_1}=-\infty$, $\mathrm{upp}_{\eta_4}=\infty$, while all other hyperparameters
are defined by the user on the base of the platform (see Section~\ref
{sec:sim}).

Figure~\ref{figHM} summarizes the full hierarchical formulation of our model.

\subsection{Choice of the \texorpdfstring{$\alpha$}{alpha} parameter}
\begin{figure}

\includegraphics{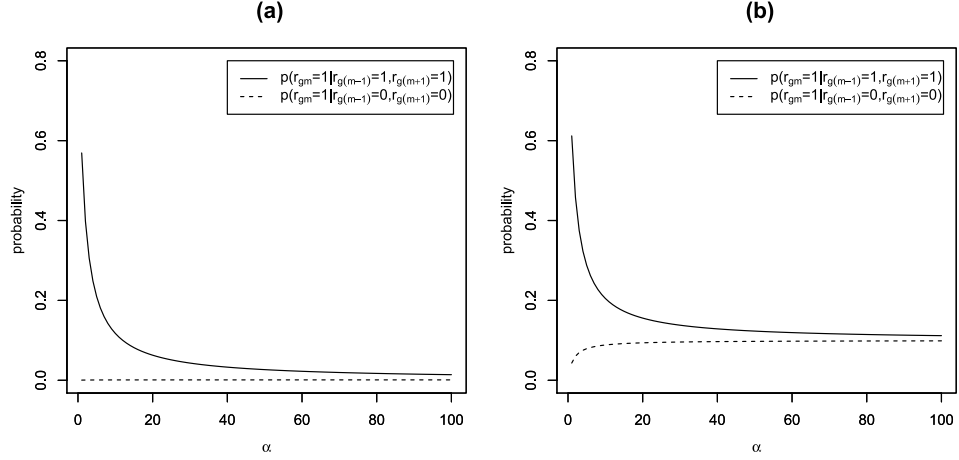}

\caption{Effect of $\alpha$ on the prior probabilities of
inclusion (\protect\ref{eq:prob_r}) for \textup{(a)} $\pi_1=0.001$
and \textup{(b)}
$\pi_1=0.1$.}
\label{fig:phase}
\end{figure}

The parameter $\alpha$ in (\ref{eq:gammaomega}) captures the relative
strength of the dependence. In particular, $\alpha=0$ implies $\gamma
_m=0$ (for $m=1,\ldots,M$), whereas $\alpha\rightarrow\infty$ leads to
$\gamma_m=1$, that is, the independent prior. In our applications, we
found that a poor choice of $\alpha$ can have undesirable effects on
the prior probability. To elucidate this further, let us arbitrarily
fix $s_{(m-1)m}=s_{m(m+1)}=0.65$. Figure~\ref{fig:phase} shows plots of
the prior probabilities (\ref{eq:prob_r}) for a grid of values of
$\alpha$ in $[1,100]$, for $\pi_1=0.001$ and $\pi_1=0.1$. These plots
show that strong dependence assumptions, that is, relatively low values
of $\alpha$, may have a differential effect on the probabilities, at
the expense of model sparsity. We notice also that the effect of
$\alpha$ is stronger when the probability of success of the Bernoulli prior is
lower. We discuss sensitivity to $\alpha$ in the simulation studies below.

%

\section{Posterior inference}
\label{sec:pinf}
Our primary interest lies in the estimation of the association matrix
$\mathbf{R}$ and the matrix of copy number states $\bolds{\xi}$. Given
that the
posterior distribution is not available in closed form, we design a
Markov chain Monte Carlo algorithm, based on stochastic search variable
selection algorithms. Once we integrate out $\bolds{\mu}$, $\bolds
{\beta}_g$
and $\sigma^2_g$, the marginal likelihood reduces to
%
%
\begin{equation}\quad
\label{eq:likelihood} f(\Yg|\bolds{\xi},\R)=\frac{(2\pi)^{-{n}/{2}}
({\cmu}/{(\cmu+n)})^{{1}/{2}}
(\cb)^{{k_g}/{2}}\Gamma({(n+\delta)}/{2})
({d}/{2})^{{\delta}/{2}}}{ \vert\Ug\vert^{
{1}/{2}}\Gamma
({\delta}/{2})({(d+q_g)}/{2})^{({(n+\delta)}/{2})}},
\end{equation}
where $q_g=\mathbf{Y}'_g\Hs\Yg- \mathbf{Y}'_g\Hs\xiR\Ug^{-1}
\xiRT\Hs\Yg$, $\Ug
=\cb\I_{k_g}+\xiRT\Hs\xiR$ and $\Hs=\I_n-\frac{\mathbf
{1}_n\mathbf{1}'_n}{n+\cmu}$, with $k_g$ indicating the number of selected
regressors for the $g$th regression function. We give full details of
our MCMC algorithm in the supplementary material [\citet{supplmat}]. The
updates at a generic iteration can be described as follows:

\begin{itemize}
\item Update $\R$ via a Metropolis step. We first select $n_g$
genes at random using a geometric distribution with parameter $p_R$.
Then, for each selected gene, with probability $\rho$ we choose between
an Add/Delete or Swap move; for the Add/Delete move we select at random
one of the elements in the corresponding row of $\R$ and change its
value (from $0$ to $1$, or vice versa); for the Swap move we select two
elements with different inclusion status and swap their values. In
updating~$\R$, we do not consider CGH probes called in copy neutral
state in more than $n\times p_{\mathrm{MC}}$ samples at the current MCMC
iteration (with $p_{\mathrm{MC}}$ set by the user), since these would not be
expected to be associated with changes in mRNA transcript abundance.
The proposed move is then accepted with probability
\[
 \min\biggl[ \frac{f(\mathbf{Y}|\bolds
{\xi},\R^{\mathrm{new}})\pi(\R
^{\mathrm{new}}|\bolds{\xi})}{f(\mathbf{Y}|\bolds{\xi},\R^{\mathrm{old}})\pi(\R
^{\mathrm{old}}|\bolds{\xi
})},1
\biggr].
\]
Since all moves are symmetric, the proposal distribution does not
appear in the previous ratio.

\item Update $\bolds{\xi}$ via a Metropolis step. This step consists
of choosing at random a column of $\bolds{\xi}$, say, $m$, and updating
the values of $n_m$ of its elements, selected at random using a
geometric distribution with parameter $p_{\xi}$. For each element, a
candidate state is sampled using the current transition matrix $\A$
[i.e., we propose $\xi_{im}^{\mathrm{new}}$ based on $\xi_{i(m-1)}^{\mathrm{old}}$] and
the proposal is accepted with probability
\[
\min\biggl[ \frac{f(\mathbf{Y}|\bolds
{\xi} ^{\mathrm{new}},\R) f(\mathbf{X}|\bolds{\xi
}^{\mathrm{new}}) \pi(\R|\bolds{\xi} ^{\mathrm{new}}) \pi(\bolds{\xi}
^{\mathrm{new}}|\bolds{\xi
}^{\mathrm{old}},\A) q(\bolds{\xi} ^{\mathrm{old}} | \bolds{\xi} ^{\mathrm{new}})}{f(\mathbf
{Y}|\bolds{\xi}
^{\mathrm{old}},\R) f(\mathbf{X}|\bolds{\xi}^{\mathrm{old}}) \pi(\R|\bolds{\xi}
^{\mathrm{old}}) \pi
(\bolds
{\xi} ^{\mathrm{old}}|\bolds{\xi}^{\mathrm{old}},\A) q(\bolds{\xi} ^{\mathrm{new}} | \bolds
{\xi} ^{\mathrm{old}})},1
\biggr].
\]

\item Update $\eta_j$, for $j=1,\ldots,4$, via a Gibbs step. We sample
$\eta_j|\mathbf{X},\bolds{\xi},\sigma_j \sim N(\n_j,\theta
_j^{-2}) \I
_{\{
\mathrm{low}_{\eta_j}<\eta_j<\mathrm{upp}_{\eta_j}\}}$, with precisions $\theta
_j=\tau
_j^{-2}+n_j \sigma_j^{-2}$ and weighted means $\n_j=\theta
_j^{-2}(\delta
_{j}\tau_j^{-2}+\Xbar_j n_j \sigma_j^{-2})$, with $n_j=\sum
_{m=1}^{M}\sum_{i=1}^{n} \I_{\{ \xi_{im}=j \}}$ and $\Xbar_j=\frac
{1}{n_j}\sum_{m=1}^{M}\sum_{i=1}^{n} X_{im} \I_{\{ \xi_{im}=j \}}$.

\item Update $\sigma_j$, for $j=1,\ldots,4$, via a Gibbs step. We
sample $\sigma_j|\mathbf{X},\bolds{\xi},\eta_j \sim \operatorname{IG} (b_j+\frac
{n_j}{2},l_j+\frac{V_j}{2}) \I_{\{\sigma_j^{-2}>\mathrm{upp}_{\sigma_j}\}}$,
where $n_j=\sum_{m=1}^{M}\sum_{i=1}^{n} \I_{\{ \xi_{im}=j \}}$ and
$V_j=(X_{im} - \eta_j)^2 \I_{\{ \xi_{im}=j \}}$.

\item Update $\A$ via a Metropolis step. We generate a new vector
for each row of $\A$ as $\A_{.j}^{\mathrm{new}}|\bolds{\xi} \sim
\operatorname{Dir}(\phi_1+\mun,\phi_2+\md,\phi_3+\mt,\phi_4+\mq)$, where $o_{hj}=\sum
_{i=1}^{n}\sum_{m=1}^{M-1} \I_{\{ \xi_{im}=h,\xi_{i(m+1)}=j \}}$,
and accept it with
probability
\[
\min\Biggl[ 1, \prod_{i=1}^{n}
\frac{\pi_{A^{\mathrm{new}}}(\xi
_{i1})}{\pi
_{A^{\mathrm{old}}}(\xi_{i1})} \Biggr].
\]
\end{itemize}

Given the MCMC output, we first perform inference on ${\mathbf
R}$ by calculating the marginal posterior probability of inclusion
(PPI) for each element, estimated by counting the number of iterations
that element was set to 1, after burn-in. A~selection is then made by
looking at those elements of $\mathbf{R}$ that have marginal PPI greater
than a value that guarantees an expected rate of false detection
(Bayesian FDR) smaller
than a fixed threshold, which we set at 0.05. We follow \citet
{Newton:2004} and calculate the Bayesian FDR as $\mathit
{FDR}_B(k)=\frac
{\sum_g \sum_m (1-\mathit{PPI}_{gm}) \I_{k}}{\sum_g \sum_m \I_{k}}$, where
$k$ is
the threshold on the PPI and $\I_{k}$ is an indicator function such
that $\I_{k}=1$ if $(1-\mathit{PPI}_{gm}) \le k$. We then estimate $\bolds{\xi
}$ by
calculating, for each position, the most frequent state value. The MCMC
output also allows us to make inference on the HMM parameters, that is,
the transition matrix $\mathbf{A}$ and the means and variances of the
emission distributions in \eqref{modelCGH}.

\section{Simulation studies}
\label{sec:sim}
We study the performance of our model on a set of simulated scenarios.
The normal human genome is diploid. However, recent studies have
reported that as much as 12\% of the human genome is variable in copy
numbers [\citet{Redon:2006}]. When copy number changes occur, they
affect segments of DNA, so neighboring chromosomal regions are expected
to have similar copy numbers. Furthermore, transitions from copy number
variants to the diploid state are expected to be more likely than
transitions between different copy number variants (e.g., from one-copy
deletion to one-copy duplication). Taking those considerations into
account, we generated a synthetic $n \times M$ matrix of copy numbers,
$\bolds{\xi}$, as follows:
\begin{itemize}
\item We initialized the matrix $\bolds{\xi}$ with all elements set
to $2$.
\item We randomly selected $L<M$ columns (including some stretches of
adjacent columns) and generated their values using the following
transition matrix:
\begin{eqnarray*}
\pmatrix{ 0.7500 &0.1800 &0.0500 &0.020\vspace*{2pt}
\cr
0.4955 &0.0020 &0.4955
&0.007\vspace*{2pt}
\cr
0.0200 &0.1800 &0.7000 &0.010\vspace*{2pt}
\cr
0.0001
&0.3028 &0.1000 &0.597 }.
\end{eqnarray*}
\item We randomly selected additional $\frac{M-L}{2}$ columns. For each
column, we generated $10\%$ of its values according to the transition
matrix above.
\end{itemize}
Following \citet{Guha:2008}, we sampled the copy number state for the
first CGH probe from the initial probability vector $\pi_{\A}$,
obtained as the normalized left eigenvector associated with the
eigenvalue 1. Given the resulting states, we generated the matrix
$\mathbf{X}$ as in \eqref{modelCGH}, where we fixed $\eta_1=-0.65,
\eta_2=0,
\eta
_3=0.65, \eta_4=1.5$ and $\sigma_1=0.1, \sigma_2=0.1, \sigma_3=0.1,
\sigma
_4=0.2$. We simulated the association network $\R$ as follows. First we
set all the $M-L$ columns equal to $0$. From the remaining columns we
selected a total of $l$ elements and set those to $1$. We set all the
remaining elements to $0$. We then generated the regression
coefficients corresponding to the $l$ selected associations by sampling
from normal distributions, as $\beta\sim N(\beta_0,\sigma_0^2)$, where
$\beta_0,\sigma_0$ were fixed as detailed in the next sections and the
signs were assigned randomly. Finally, we generated the gene expression
outcomes, $Y_{ig}$ $(g=1, \ldots, G)$ as $Y_{ig}=\mu_g +\bolds{\xi
}_{i}\bolds
{\beta}_g + \varepsilon_{ig}$ with $\mu_g\sim N(0,\sigma^2_{\mu_g})$,
$\sigma_{\mu_g}=0.1$, and $\varepsilon_{ig} \sim N(0,\sigma
^2_{\varepsilon})$.
Unless otherwise specified, in the following we set $n=100$, $G=100$,
$M=1000$, $L=250$, $l=20$ and $\sigma_{\varepsilon}=0.1$. We also
considered simulated scenarios with a different $\sigma_{\varepsilon}$
value for each gene $g$ and found similar performances to those we
report below [\citet{supplmat}].

As for hyperparameter settings, those in \eqref{prior} and \eqref
{eq:prob_r} determine the amount of shrinkage in the model. We followed
the guidelines provided by \citet{sha:2004} and chose $\cb$ in the range
of variability of the data so as to control the ratio of prior to
posterior precision. Specifically, we set $\cb=10$, in all simulations.
Furthermore, we specified vague priors on the intercept term, by
setting $\cmu=10^{-6}$, and on the error variance $\sigma_g^{2}$, by
setting $\delta= 3$ and choosing $d$ such that the expected value of
$\sigma_g^2$ represents a fraction of the observed variance of the
standardized responses (5\% for the results reported here).
For all scenarios, we considered the dependent prior model (\ref
{eq:prob_r_int}) with $e=0.001$ and $f=0.999$ and assessed sensitivity
for varying $\alpha$ in \eqref{eq:gammaomega} in the set $\{
5,10,50,100,\infty\}$. The notation $\alpha=\infty$ succinctly
indicates the independent prior.
For the HMM model, similar to \citet{Guha:2008}, we set $\eta_j\sim
N(\delta_j,\tau_j^{2}) \cdot\I_{\{\mathrm{low}_{\eta_j}<\eta_j<\mathrm{upp}_{\eta
_j}\}
}$, $\sigma_j^{-2}\sim\operatorname{Ga}(b_j,l_j)\cdot\I_{\{\sigma
<\mathrm{upp}_{\sigma
_j}\}}$
with $b_j=1$, $l_j=1$, $j=1,\ldots,4$, and the other hyperparameters
specified as in Table~\ref{tab:HMM_par}. The lower bound for $\eta_4$,
$\mathrm{low}_{\eta_4}$ was set to avoid that a large number of single copy
gains be erroneously classified as multiple copy gains. The choice of
the truncation $\sigma_j^{-2}>6$ is a mild assumption, and it is
equivalent to setting $\sigma_j<0.41$. Finally, we assumed each row of
the transition matrix as independently distributed according to
$\operatorname{Dir}(1,1,1,1)$.

%
\begin{table}
\caption{Simulation study: Specification of the HMM hyperparameters}
\label{tab:HMM_par}
\begin{tabular*}{\textwidth}{@{\extracolsep{\fill}}ld{2.2}d{2.2}d{1.2}c@{}}
\hline
\textbf{HMM parameters} & \multicolumn{1}{c}{\textbf{State 1}} &
\multicolumn{1}{c}{\textbf{State 2}} & \multicolumn{1}{c}{\textbf
{State 3}} & \multicolumn{1}{c@{}}{\textbf{State 4}} \\
\hline
$\delta_j$ & -1 & 0 & 0.58 & 1\\
$\tau_j$ & 1 & 1 & 1 & 2 \\
$\mathrm{low}_{\eta_j}$ & \multicolumn{1}{c}{$-\infty$} & -0.1 & 0.1 &
\multicolumn{1}{c@{}}{$\eta_3+\sigma_3$} \\
$\mathrm{upp}_{\eta_j}$ & -0.1 & 0.1 & 0.73 & \multicolumn{1}{c@{}}{$\infty$} \\
$\mathrm{upp}_{\sigma_j}$ & 0.41 & 0.41 & 0.41 & 1\\
\hline
\end{tabular*}
\end{table}

\begin{figure}[t]
\includegraphics{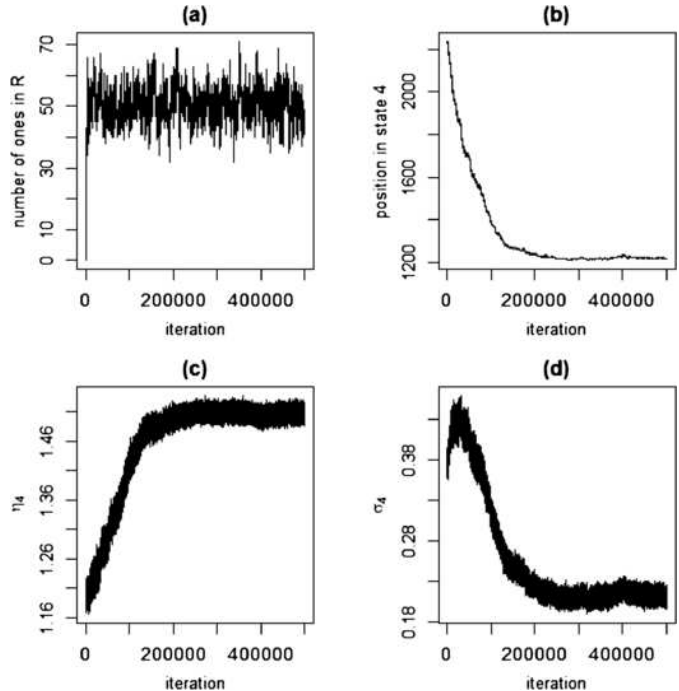}
\caption{Simulation study: Trace plots for: \textup{(a)} $\R$,
number of ones in the association matrix, \textup{(b)} $\xi_4$,
number of positions estimated as multiple gains,
\textup{(c)} $\eta_4$, mean value of the positions estimated
as multiple gains and \textup{(d)} $\sigma_4$, standard
deviation of the positions estimated as multiple
gains, for one MCMC run on simulated scenario 1.
We note that state four has the smallest number
of observations, thus, more variance and less stationarity is expected.}
\label{fig:traces}
\end{figure}

When running the MCMC chains, we sampled initial values for $\eta_j$
and $\sigma_j$ from their respective priors and initialized~$\bolds
{\xi}$
as $\xi_{im}=j$ ($j=1,\ldots,4$) if $X_{im}>T_j$ with $\mathbf{T}=[-
\infty,-0.5,0.29,0.79]$. We derived the initial value of $\A$ from the initial~$\bolds{\xi}$, based on the proportion of transitions. We set the initial
$\R$ as a matrix with all elements equal to zero. All results reported
here were obtained with MCMC chains with 500,000 iterations and a
burn-in of 350,000, fixing $p_R=0.4$, $p_\xi=0.6$, $p_{\mathrm{MC}}=0.9$ and
$\rho
=0.5$. We assessed convergence by inspecting the MCMC sample traces for
all parameters; see Figure~\ref{fig:traces} for an example of typical
plots.\vadjust{\goodbreak} Moreover, we applied the diagnostic test of \citet{geweke:1992}
for the equality of the means, based on the first $10\%$ and the last
$50\%$ of the chain. We also used the \citet{Heidel:1981} test on the
stationarity of the distribution to determine a suitable burn-in.

%

\subsection{Inference on the association network ($\mathbf{R}$)}
\label{sec:inf_R}

We present results from two simulated scenarios. The first scenario
assumes no particular (spatial) dependence structure in the association
between markers and genes. For this scenario, we generated $l$
regression coefficients as $\beta\sim N(2,0.3^2)$, except for $6$
values which we drew from $N(0.5,0.3^2)$, to take into account a lower
signal to noise ratio. In the second scenario we explicitly assumed
dependence among the regression coefficients. In particular, we
selected two clusters of adjacent CGH probes and assumed they affect
the expression of the same gene. The corresponding coefficients were
sampled as $\beta\sim N(0.5,0.3^2)$. In both scenarios, we simulated data
for two values of the error standard deviation, that is, $\sigma
_{\varepsilon}=0.1$ and $\sigma_{\varepsilon}=0.5$.

%
\begin{figure}

\includegraphics{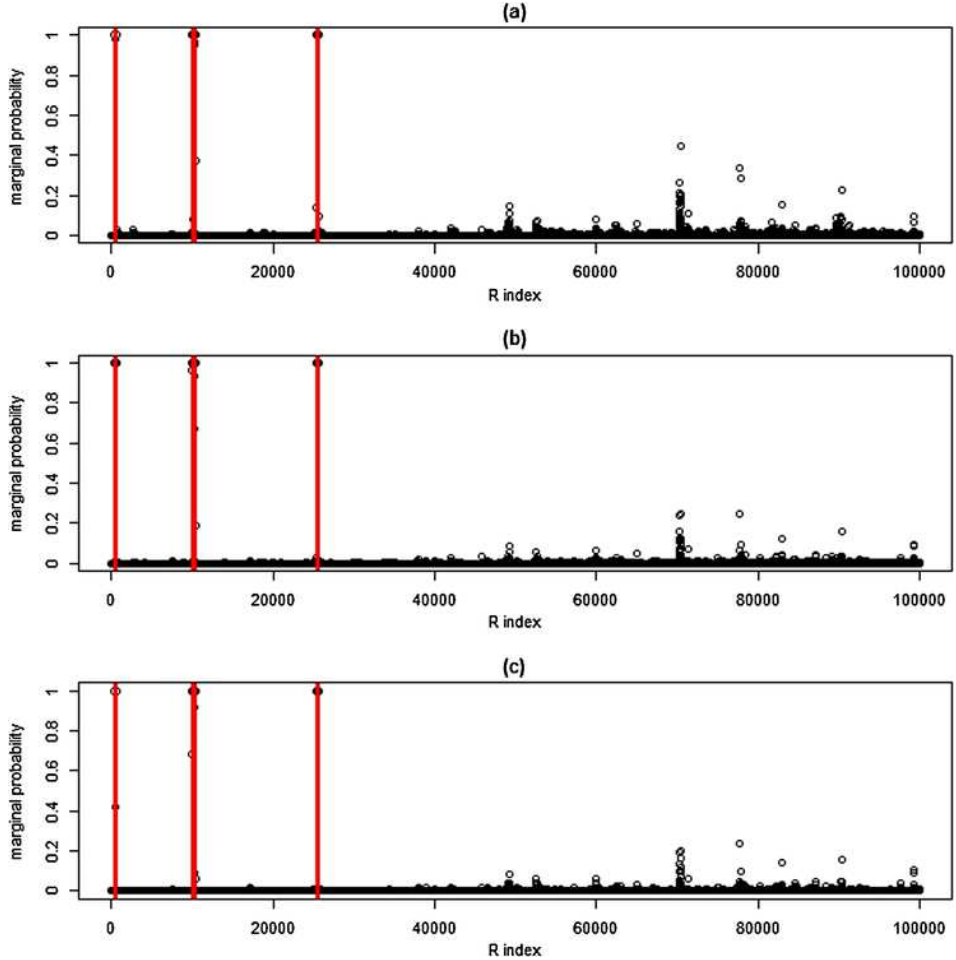}

\caption{Simulated data: Simulated scenario 1 with $\sigma
_{\varepsilon
}=0.1$:
Marginal posterior probabilities of inclusion of the elements $r_{gm}$
of the association matrix $\R$. Plots refer to prior model
(\protect\ref{eq:prob_r_int}) with \textup{(a)}~$\alpha=20$,
\textup{(b)} $\alpha
=100$, \textup{(c)}
$\alpha=\infty$ (independent prior). Vertical lines indicate the
true gene-CNVs associations.}
\label{fig:margR}
\end{figure}

We start by analyzing the results for the first scenario. Figure~\ref
{fig:margR} shows marginal PPIs of the elements $r_{gm}$ of $\R$, for
the case $\sigma_{\varepsilon}=0.1$. The model recovers well the true
gene-CNVs associations (vertical lines), although it is evident that
relatively small values of $\alpha$, implying greater a priori
dependence structure, result in an increased number of erroneous
decisions when such structure is indeed not present in the data. A
selection of the significant associations is made by considering those
elements of $\mathbf{R}$ that have marginal PPI greater than a value that
guarantees a prespecified FDR. For example, the first panel of
Table~\ref{tab:sim1R} reports results in terms of specificity, sensitivity,
false positives (FP), false negatives (FN) and number of detections,
obtained with an upper bound on the FDR set to.05.
Sensitivity is calculated as the ratio of true positive (TP) counts
over $l$ and specificity as the ratio of true negatives (TN) over
$(G\times M-l)$. In the same table we also report the realized Bayesian
$q$-value, calculated as $\min_{\{(1-\mathit{PPI}) \leq k\}} \mathit{FDR}_B(k)$;
see, for example, \citet{Broet:2004} and \citet{Morris:2004}.
Results show that a lower $\alpha$ leads to less FN calls but increased
FP counts. However, due to the large number of TNs, such effect
translates in much improved sensitivity at the expense of only a
minimal decrease in specificity. Results are similar for $\sigma
_\varepsilon=0.5$, although, as expected, the model performance improves
when the error variance is smaller (see lower panel of each scenario in
Table~\ref{tab:sim1R}).

\begin{table}
\tabcolsep=0pt
\caption{Simulated scenarios 1 and 2: Results on specificity,
sensitivity, false positives, false negatives,
number of detections and Bayesian $q$-values, for
the dependent prior model (\protect\ref{eq:prob_r_int}) and
the independent case ($\alpha=\infty$), obtained for an FDR threshold of 0.05}
\label{tab:sim1R}
\begin{tabular*}{\textwidth}{@{\extracolsep{\fill}}lcccccccc@{}}
\hline
& $\bolds{\alpha=5}$ & $\bolds{\alpha=10}$ & $\bolds{\alpha=20}$ &
$\bolds{\alpha=30}$
& $\bolds{\alpha=40}$ & $\bolds{\alpha=50}$ & $\bolds{\alpha
=100}$ & $\bolds{\alpha=\infty}$\\
\hline
\textit{Scenario} 1& \multicolumn{8}{c}{$\sigma_{\varepsilon}=0.1$} \\
Spec. &  0.99785  &  0.99795 &  0.99999  &  1  &  0.99999  &  0.99999  &  1  &  1  \\
Sens. &  0.95  &  0.95 &  0.9  &  0.95  &  0.9  &  0.9  &  0.9  &  0.8  \\
FP/FN & $215/1$ & $205/1$ & $1/2$ & $0/1$ & $1/2$ & $1/2$ & $0/2$ & $0/4$ \\
\# detect & 234 & 224 & 19 & 19 & 19 & 19 & 18 & 16 \\
$q$-value &0.048679 & 0.046491 & 0.03444 & 0.042294 & 0.045403 & 0.048651 & 0.024107 & 0.024674 \\[3pt]
&\multicolumn{8}{c}{$\sigma_{\varepsilon}=0.5$}\\
Spec. &  0.99999  &  0.99999 &  0.99999  &  1  &  1  &  1  &  0.99999  &  0.99999  \\
Sens. &  0.95  &  0.95 &  0.9  &  0.9  &  0.9  &  0.85  &  0.8  &  0.8  \\
FP/FN & $10/1$ & $1/1$ & $1/2$ & $0/2$ & $0/2$ & $0/3$ & $1/4$ & $1/4$ \\
\# detect & 29 & 20 & 19 & 18 & 18 & 17 & 17 & 17 \\
$q$-value & 0.046464 & 0.041118 & 0.049538 & 0.038603 & 0.0428 & 0.026924 & 0.028897 & 0.033866 \\[6pt]
\textit{Scenario} 2 &\multicolumn{8}{c}{$\sigma_{\varepsilon}=0.1$} \\
Spec. &  0.99987  &  0.99998 &  0.99999  &  0.99999  &  0.99999  &  0.99999  &  0.99999  &  0.99999  \\
Sens. &  0.95  &  0.95 &  0.95  &  0.95  &  0.95  &  0.95  &  0.9  &  0.85  \\
FP/FN & $13/1$ & $2/1$ & $1/1$ & $1/1$ & $1/1$ & $1/1$ & $1/2$ & $1/3$ \\
\# detect & 32 & 21 & 20 & 20 & 20 & 20 & 19 & 18 \\
$q$-value & 0.045476 & 0.0452311 & 0.031514 & 0.042635 & 0.044119 & 0.046781 & 0.04567 & 0.035927 \\[3pt]
&\multicolumn{8}{c}{$\sigma_{\varepsilon}=0.5$}\\
Spec. &  0.99989  & 0.99994  &  0.99998  &  0.99998  &  0.99998  &  0.99998  &  0.99998  &  0.99998  \\
Sens. & 0.85   & 0.85  &  0.85  &  0.8  &  0.8  &  0.8  &  0.7  &  0.6  \\
FP/FN & $11/3$ & $6/3$ & $2/3$ & $2/4$ & $2/4$ & $2/4$ & $2/6$ & $2/8$ \\
\# detect & 28 & 23 & 19 & 18 & 18 & 18 & 16 & 14 \\
$q$-value & 0.04506 & 0.049371 & 0.039412 & 0.041290 & 0.045759 & 0.047261 & 0.047235 & 0.047865 \\
\hline
\end{tabular*}
\end{table}

\begin{figure}

\includegraphics{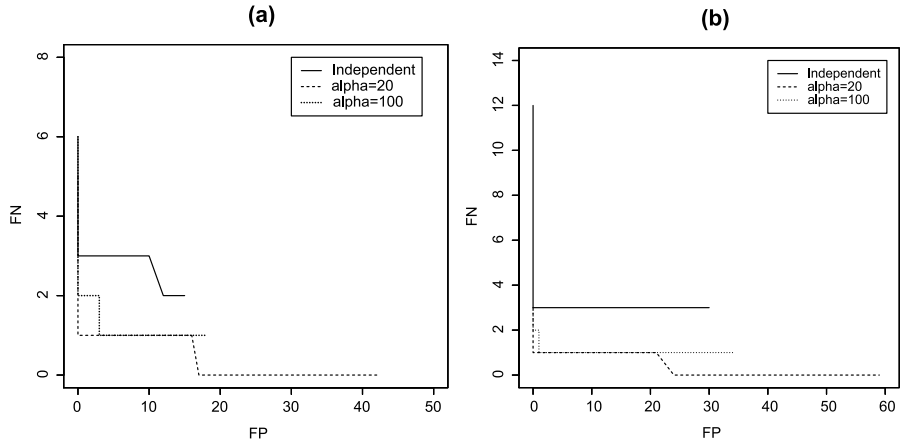}

\caption{Simulated scenarios 1\textup{(a)} and 2\textup{(b)} with
$\sigma_{\varepsilon}=0.1$:
Numbers of FP and FN obtained by considering different thresholds on
the marginal probabilities of inclusion of Figure \protect\ref{fig:margR}.
Threshold values are calculated as a grid of equispaced points in the
range $[0.07,1]$. Plots refer to prior model~(\protect\ref
{eq:prob_r_int}) with
different values of $\alpha$.}\label{fig:ROC}
\end{figure}

In order to investigate the effect of the threshold on the PPIs on the
selection results, in Figure~\ref{fig:ROC}{(a)} we report ROC-type
curves displaying FP counts versus FN counts calculated at a grid of
equispaced thresholds in the interval $[0.07,1]$. The plots clearly show
that dependent priors obtained for lower values of $\alpha$ generally
outperform the independent case, regardless of the threshold.\looseness=1

%

Our results are confirmed by the second simulated scenario. As
expected, dependent priors improve the FP counts (see the last two
panels of Table~\ref{tab:sim1R}), since the spatial dependence in the
gene-CNVs association structure is now explicitly taken into account.
Indeed, the independent prior shows
worse performance, due to its inability to use information gathered
from adjacent probes. As in the first simulated scenario, we again
notice that lower values of $\alpha$ lead to less FN calls but
increased FP counts; see Table~\ref{tab:sim1R} and Figure~\ref
{fig:ROC}{(b)}. As a general guideline regarding the choice of this
parameter, our results indicate that moderate values of $\alpha$ give
an appropriate compromise between false positives and false negatives.
See Section~\ref{sec:dis} for additional discussion.

\subsection{Inference on the CNV states (\texorpdfstring{$\bolds{\xi}$}{xi}) and the HMM
parameters}
\label{sec:inf_xi}

We now turn to the inference on the CGH states, encoded by the matrix
$\bolds{\xi}$. Table~\ref{tab:sim1xi} reports the misclassification counts
and corresponding percent rates. In order to compute these summary
statistics, for each element we considered the modal state attained at
each genomic location over all MCMC iterations (after burn-in). The
misclassification rates appear to be consistent over the different
values of $\alpha$ and of the error standard deviation $\sigma
_{\varepsilon
}$. A close look at the distribution of the misclassifications over the
four states showed that most errors occur between adjacent classes
(results not shown).

%
\begin{table}[b]
\tabcolsep=0pt
\caption{Simulated scenarios 1 and 2: Results on $\bm\xi$ as number of misclassified
copy number states, for the dependent prior model (\protect\ref{eq:prob_r_int})
and various values of $\alpha$}\label{tab:sim1xi}
\begin{tabular*}{\textwidth}{@{\extracolsep{\fill}}lcccccccc@{}}
\hline
{\textbf{\# Miscl.}} & & & & & & & &\\
{\textbf{(percent)}} & $\bolds{\alpha=5}$ & $\bolds{\alpha=10}$ &
$\bolds{\alpha=20}$ & $\bolds{\alpha=30}$ &
$\bolds{\alpha=40}$ & $\bolds{\alpha=50}$ & $\bolds{\alpha=100}$
& $\bolds{\alpha=\infty}$\\
\hline
Scenario 1 & 179 & 162 & 78 & 78 & 77 & 74 & 74 & 78\\
$\sigma_{\varepsilon}=0.1$ & $(0.179\%)$ & $(0.162\%)$ & $(0.078\%)$ & $(0.078\%)$ & $(0.077\%)$ & $(0.074\%)$ & $(0.074\%)$ & $(0.078\%)$\\
Scenario 1 & \phantom{0}68 & \phantom{0}71 & 70 & 69 & 76 & 68 & 72 & 73\\
$\sigma_{\varepsilon}=0.5$ & $(0.068\%)$ & $(0.071\%)$ & $(0.07\%)$ & $(0.069\%)$ & $(0.076\%)$ & $(0.068\%)$ & $(0.072\%)$ & $(0.073\%)$\\
Scenario 2 & \phantom{0}51 & \phantom{0}58 & 62 & 53 & 60 & 61 & 60 & 62\\
$\sigma_{\varepsilon}=0.1$ & $(0.051\%)$ & $(0.058\%)$ & $(0.062\%)$ & $(0.053\%)$ & $(0.06\%)$ & $(0.061\%)$ & $(0.06\%)$ & $(0.062\%)$\\
Scenario 2 & \phantom{0}60 & \phantom{0}59 & 60 & 55 & 60 & 53 & 53 & 54\\
$\sigma_{\varepsilon}=0.5$ & $(0.06\%)$ & $(0.059\%)$ & $(0.06\%)$ & $(0.055\%)$ & $(0.06\%)$ & $(0.053\%)$ & $(0.053\%)$ & $(0.054\%)$\\
\hline
\end{tabular*}
\end{table}


Our model allows also to conduct inference on the parameters of the
HMM, that is, the transition matrix $\mathbf{A}$ and the means and
variances of the emission distributions in model \eqref{modelCGH}. As
an example, scenario 1 ($\sigma_{\varepsilon}=0.1$) using the independent
prior gave the following estimates:
$\hat{\bolds{\eta}}=[-0.64963,0.00044,\break 0.64936, 1.50717]$ and
$\hat{\bolds{\sigma}}=[0.10206,0.09994,0.10069,0.21187]$, which appear to be all very
close to
the simulated values, with the exception of $\sigma_4$ which is
slightly overestimated. This is the standard deviation of the
amplification state, that collects all copy number gains larger than 1,
so some overestimation might be expected. We obtained similar results
in all other simulations we considered. As for the transition matrix
across CGH states, the estimates appeared close to the truth [result
reported in \citet{supplmat}].

\subsection{Comparison with single stage approaches}
We compare the results based on our unified method, which performs
simultaneous CNV detection and selection of significant associations,
to single stage approaches that focus solely on CNV detection or solely
on association analysis using the raw measurements.

Using the CNV detection method of \citet{Guha:2008}, which analyzes each
sample separately, and specifying the same prior settings as our model,
there were, respectively, $2695$ and $8349$ misclassified CNV calls for
the two scenarios with $\sigma_{\varepsilon}=0.1$ (instead of $78$
and $62$
as reported for the independent case in Table~\ref{tab:sim1xi}). This
result demonstrates that the integration of multiple samples and the
joint modeling of gene expression data offer improved estimation of
copy number states.

We also looked into the performance of Bayesian variable selection in a
regression model where the predictors are the raw continuous CGH
measurements, therefore ignoring the inference of the latent copy
number states. For the prior on the variable selection indicators,
since the copy number states were not estimated, we cannot use prior
model (\ref{eq:prob_r}). Instead, we assumed the independent prior
$r_{gm}\sim \operatorname{Bern}(\pi_1)$ and set $\pi_1=0.001$. For $\sigma
_{\varepsilon
}=0.1$, using an FRD threshold of 0.05, we obtained $\operatorname
{specificity}=1$ and $\operatorname{sensitivity}=0.7$ in the first
simulated scenario and $\operatorname{specificity}=1$ and
$\operatorname
{sensitivity}=0.2$ in the second scenario.
In both cases the performance of the competing model worse than to that
of our model with the independent prior (see Table~\ref{tab:sim1R}). In
particular, in the second scenario the model with the dependent prior
outperforms both the model with the independent prior and the competing
model that uses the raw continuous CGH measurements.

\section{Case study on human cancer cell lines}
\label{sec:case}
We applied our model to the analysis of the NCI-60 cell line panel,
which consists of 60 human cancer cell lines derived from a diverse set
of tissues (brain, bone marrow, breast, colon, kidney, lung, ovary,
prostate and skin). We downloaded the normalized aCGH Agilent 44K data
and the Affymetrix HG-U133A RMA gene expressions using CellMiner
(\surl{discover.nci.nih.gov/cellminer}).
In the current analysis, we excluded cell line $40$ from the data set,
since no gene expression measurements were available in the repository.
We imputed the remaining missing values using the $k$-nearest neighbor
algorithm with $k=5$.

In performing our analysis we employed pathway-based scores of the gene
expression data. This strategy helped us to reduce the dependence
between the outcome variables in model \eqref{systemeqs} and also to
achieve a dimension reduction of the model space. Methods that employ
pathway-based scores of gene expression data have become quite popular
in genomics; see, for example, \citet
{Su:2009,Ovacik:2010,Chen:2010,Drier:2013}, among others. More
precisely, we considered the genes that map to each one of the 186 KEGG
pathways, using the software Compadre [see \citet{Trevino:2012}]. Then,
for each pathway, we applied principal component analysis (PCA) to the
gene expression data and selected the components that explained at
least $80\%$ of the variability. This procedure led us to the selection
of $G=3195$ pathway components, which we used as response variables in
model~\eqref{systemeqs}. Furthermore, we considered the 1521 CGH probes
mapping to chromosome 8 and selected those that showed variability
across tissue types via an ANOVA test with multiplicity correction.
This resulted in a set of $M = 89$ CHG predictors.

For model fitting, we used hyperparameter settings similar to those
used in the simulation scenarios described in Section~\ref{sec:sim}. We
ran $100\mbox{,}000$ iterations with a burn-in of $50\mbox{,}000$, setting
$p_R=0.1$,
$p_{\xi}=0.3$ and $p_{\mathrm{MC}}=0.9$ in the MH proposals.
As suggested by the results of the simulations, we set $\alpha$ to a
relatively small value, that is, $\alpha=25$. For comparisons, we also
looked at the case $\alpha\rightarrow\infty$ (i.e., the
independent prior). As in the simulation study, we assessed convergence
by inspecting the MCMC sample traces for all parameters. Moreover, we
applied the Geweke diagnostic test for the equality of the means and
the Heidelberger and Welch test on the stationarity of the distribution
to determine a suitable burn-in.

\begin{figure}

\includegraphics{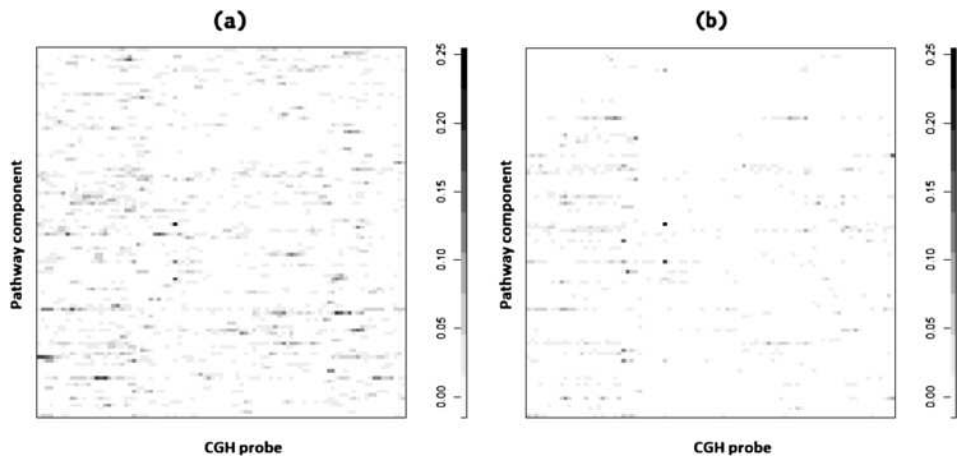}

\caption{Case study: Heatmaps of PPIs of pathway-CNV associations using
the dependent prior with $\alpha=25$ \textup{(a)} and the independent prior
\textup{(b)}.}\label{fig:Hea}
\end{figure}

We ranked the marginal PPIs of the elements of $\mathbf{R}$ in order to
identify the most significant associations. Figure~\ref
{fig:Hea}{(a)} shows a heatmap of the pathway-CNV associations with highest PPI
for the case $\alpha=25$ (roughly the top 100 associations, which
correspond to a threshold of $0.07$ on the PPIs). Figure~\ref
{fig:Hea}{(b)} shows the same selection for the independent prior.
Notice that the latter heatmap is more sparse. In addition, the heatmap
for $\alpha=25$ shows a stronger tendency to include groups of adjacent
CGH probes as significant for the same pathway component, which is
coherent with how we built our prior probability model.

%

\begin{figure}

\includegraphics{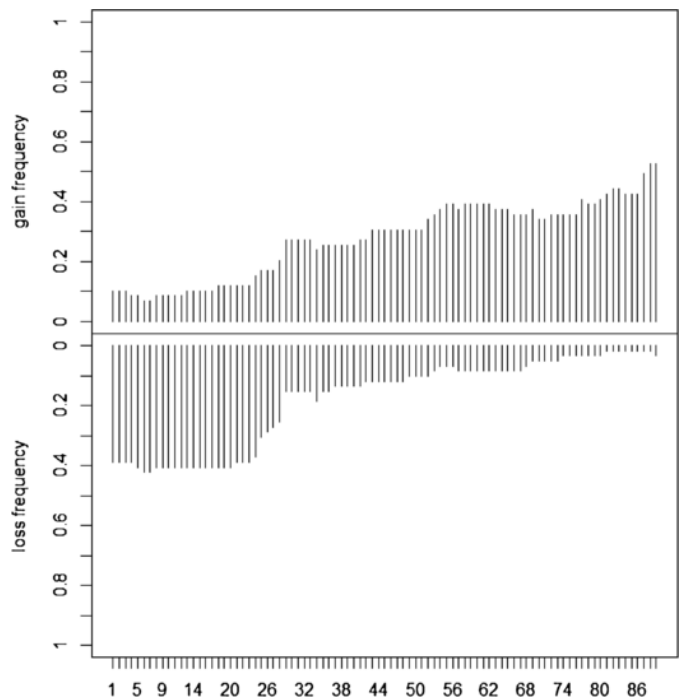}

\caption{Case study: Proportion of estimated gains and losses among the
59 samples for the 89 CGH probes considered.}
\label{fig:xi}
\end{figure}

As for inference on the copy number states, the estimates of the state
specific means and variances were $[-0.6419,-0.0105,0.49,1.0236]$ and
$[0.2059,\break 0.08115,0.1287,0.27138]$, respectively, which are consistent with
the theoretical values. Furthermore, the estimated transition matrix
well captured the state persistence of the CGHs (results not shown). We
also notice that the first and the last value of the vector of
estimated variances are larger than those corresponding to neutral and
single gain states. This is what we would expect, since the first and
the last class correspond to multiple copy number losses and gains,
respectively. Finally, Figure~\ref{fig:xi} shows the estimated
frequencies of gains (single and multiple) and losses plotted along the
samples for each of the 89 CGH probes considered for analysis.

%

\subsection{Biological interpretation of our findings}

\begin{figure}

\includegraphics{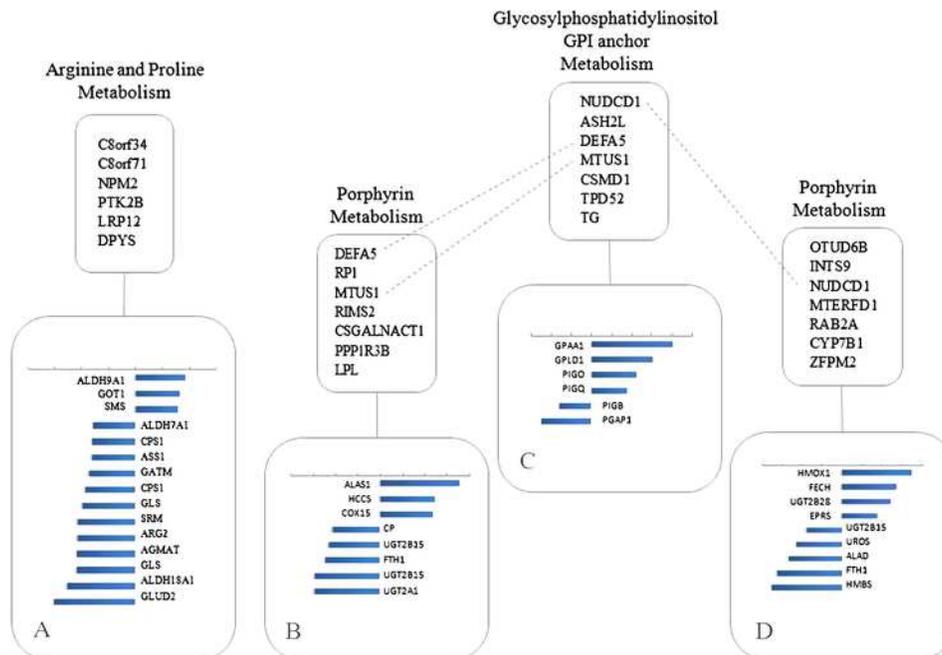}

\caption{Case study: Schematic representation of selected
associations, showing the selected genes with CNVs and the
transcriptional predicted target genes (top $20\%$ of the absolute
value of the PC loadings). Bars represent the PC loading values. Dashed
lines point at genes with CNVs that overlap across pathways.}
\label{fig:fal1}
\end{figure}

Our results identify potential links between genomic mutations, in the
form of CNVs, and the transcriptional activity of target pathways. In
this section we explore the biological significance of the identified
associations and assess whether they can be used to generate
biologically relevant hypotheses. Figure~\ref{fig:fal1} is a schematic
representation of the conceptual relationships between genes linked to
CGHs for a set of 4 target pathway components. The 4 pathway components
were selected as those with the highest numbers of associations in
Figure~\ref{fig:Hea}. For each of the 4 components we report the top
$20\%$ of the genes with highest PC loadings [subplots (A), (B), (C), (D), with
bars representing the PC loading values] as those with highest
expression variability.
Selected genes with CNVs are also listed below the pathway names.
Finally, dashed lines point at genes with CNVs that overlap across
selected pathways. These results identify two main molecular pathway
blocks. The first [Figure~\ref{fig:fal1}(A)] represents the connection
between six genetic mutations with Arginine metabolism. The second
[Figure~\ref{fig:fal1}(B), (C), (D)] represents a partially overlapping set of 18
genomic mutations and the expression of genes involved in
Glycosylphosphatidylinositol (GPI) anchor metabolism and Porphyrin metabolism.

%

There is strong evidence linking Arginine metabolism to cancer in the
literature. For example, arginine metyiltransferases are key enzymes in
modulating DNA methylation, a primary mechanism in neoplastic
transformation [\citet{Yang:2013}]. A connection between Arginine
metabolism and suppressor cells in cancer has also been proposed
[\citet
{Raber:2012}]. Our results therefore suggest that the expression of a
number of enzymes involved in Arginine metabolism may be linked to
specific mutations. Interestingly, several of these mutations are known
cancer genes. For example,
it has been shown that mutations in Nucleoplasmin 2 (NPM2), a core
histone chaperone involved in chromatin reprogramming, are associated
to increase resistance in a cancer cell line [\citet{Dalenc:2012}]. In
the supplementary material [\citet{supplmat}] we report details of the
functions of other mutations linked to the target pathways we have identified.

Our results also identify a partially overlapping set of mutations
linked to GPI-anchor metabolism and Porphyrin metabolism [Figure~\ref
{fig:fal1}(B), (C), (D)]. Similar to the Arginine metabolism, over-expression
of several enzymes in the GPI-anchor metabolism has been shown to
induce tumorigenesis and invasion in human breast cancer [\citet
{Wu:2006}]. On the other hand, no direct link between the expression of
Porphyrin metabolism genes and cancer has been reported, although there
is evidence that increased porphyrins may be a parallel disease in
liver cancer models [\citet{Kaczynski:2009}].

\begin{figure}[t]
\includegraphics{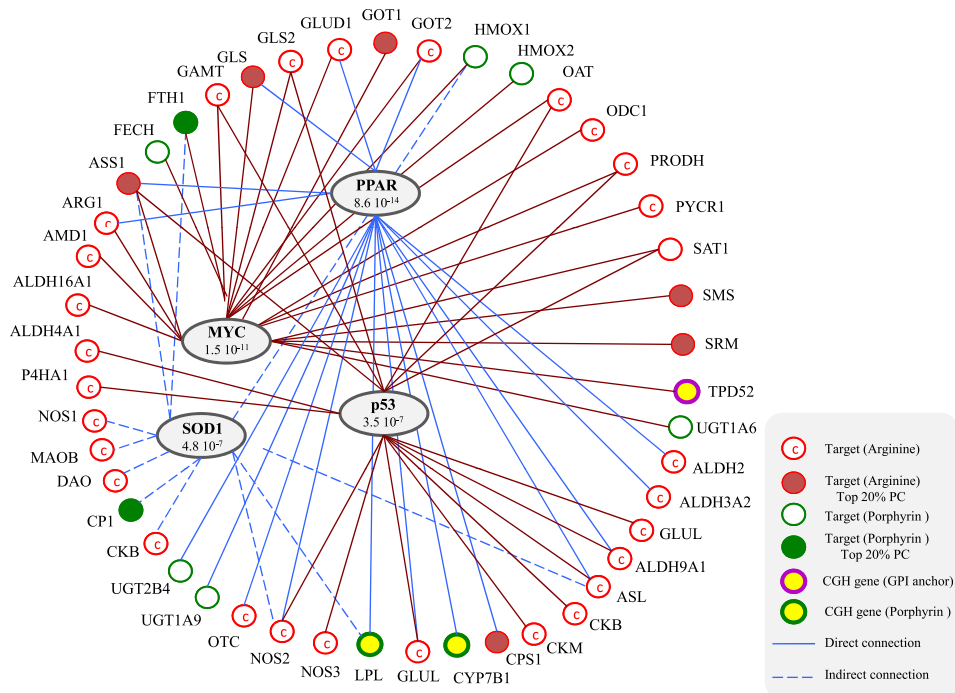}
\caption{Case study: Potential upstream regulators of the selected
genes with CNVs and target genes. The plot shows the top 4 most likely
regulators $(p<10^{-7}$), that is, PPAR, the oncogenes MYC and p53 and
the ROS scavenger SOD1. These are upstream to many of the Arginine
metabolism genes (represented by the red circles), including a large
number of those in the top 20\% of the PC loadings (filled red
circles). Some Porphyrin metabolism transcriptional targets are also
included (green circles and filled green circles). Furthermore, 3 of
the selected genes with CNVs are linked to the 4 regulators (yellow
filled green and red circles).}
\label{fig:fal2}
\end{figure}

Having identified possible relationships between genomic mutations and
target functional pathways, we wondered whether these might be also
linked to already known regulators involved in cancer. To test this
hypothesis, we looked at whether the lists of genes identified either
as genetic mutations or target genes are enriched for targets of known
regulators. More specifically, we searched for putative (directed or
indirect) upstream regulators of all genes involved in the Arginine,
GPI-anchor and Porphyrin metabolisms as well as putative upstream
regulators of the genes with CNVs selected by the model. We searched a
database of known targets of transcription factors and other regulators
(\surl{www.ingenuity.com}) and used a Fisher's exact test to assess
whether there was a statistically significant overlap between the genes
in our lists and the genes regulated by each regulator in the database.
In this analysis we used a high stringency threshold ($p <10^{-6}$) to
define putative regulators.
Figure~\ref{fig:fal2} shows our findings. All 4 putative upstream
regulators identified at the high stringency threshold were genes known
to be of primary importance in cancer biology. These were the
well-known oncogenes MYC and p53, the Peroxisome proliferator-activated
receptor PPAR [\citet{Belfiore:2009}] and the reactive oxygen species
scavenger Superoxide dismutase SOD1 [\citet{Sommar:2011} and \citet
{Noor:2002}]. We found that genes connected to these regulators were
primarily representing enzymes involved in Arginine metabolism ($76\%$
of the total targets, $35/46$) representing $50\%$ ($35/72$) of the genes
in that pathway. Of these, 6 represented genes within the top $20\%$ PC
loadings [Figure~\ref{fig:fal1}(A)]. Eight genes connected to the 4
identified regulators ($17\%$ of the total targets) were representing
enzymes in the Porphyrin metabolism pathway (representing $17\%$ of the
total pathway genes, $8/46$). Interestingly, no genes with CNVs selected
in the Arginine metabolism model were linked to the 4 regulators.
Instead, 2 of the 3 genes with CNVs included in Figure~\ref{fig:fal2}
were in the Porphyrin metabolism pathway block and 1 in the GPI-anchor
metabolism. Overall, these findings support the hypothesis that the
associations we have identified represent genes highly implicated in cancer.

\section{Discussion}
\label{sec:dis}
In this paper we have developed a hierarchical Bayesian modeling
framework for the integration of high-throughput data from different
sources. We have focused in particular on gene expression levels and
CGH array measurements, collected on the same subjects. Our modeling
framework has several innovative features. First, it allows the
identification of the joint effects of multiple CNVs on mRNA transcript
abundance. Second, it reduces the bias that arises when ignoring the
uncertainty in the CNV estimation process (i.e., using copy number
calls as if they were the true states),
by allowing the simultaneous inference of CNVs and their association to
gene expression. We have shown in simulations that noise in the raw
measurements leads to the detection of spurious associations and also
that it is advantageous to incorporate the estimation of copy numbers
into the analysis, as this reduces the detection of false positive
associations. Findings from an analysis we have conducted on data from
60 cancer cell lines support the hypothesis that the model we have
developed has the potential to identify important linkages between gene
expression and CNVs. The data set we have considered spans a large
spectrum of tissues and cancer types.
It is expected that the detection power of our approach will be higher
with more defined patient populations. These studies will require
dedicated clinical studies.

%

Our model aims to identify contiguous regions of DNA aberration that
jointly affect the expression of a gene. To accomplish this, we have
specified selection priors that cleverly account for spatial dependence
across DNA segments. This prior model depends on a parameter, $\alpha$,
that plays an important role in capturing the dependence structure. We
investigated the option of putting a prior distribution on this
parameter. However, with a Gamma prior, for example, and a
Metropolis--Hastings step to sample $\alpha$, the data only have an
indirect effect on the MH acceptance ratio, via the definition of
$s_{(m-1)m}$ and the values of $r_{gm}$, and the MH ratio is dominated
by the prior probability of $r_{gm}$. As seen in Figure~\ref
{fig:phase}, the prior probability of inclusion/exclusion increases if
the neighbors are included/excluded, and this effect is particularly
dramatic for the prior probability of inclusion under lower values of
$\alpha$. This causes the sampler to move to regions of the posterior
characterized by higher dependence between contiguous states, accepting
a move every
time a smaller value of $\alpha$ is proposed. Such behavior could be
prevented by introducing a second parameter in the prior, in order to
penalize for large numbers of included links. The construction of such
prior will need further investigation on our part. We find the
single-parameter prior model we have proposed here rather intuitive and
easy to specify. In our simulations we have found values of $\alpha$ in
the range $\alpha=[20, 50]$ to work well, leading to a good balance
between the number of FN and FP. Results shown in Table~\ref
{tab:sim1R}, in fact, are clearly robust to the choice of $\alpha$ in
this range. Values lower than 20 lead to a steady increase in the
number of included links, while values higher than 50 result in priors
closer and closer to the independent model. Moreover, for all the
simulated examples and all $\alpha$ values in the suggested range, the
top 15 links identified
with highest posterior probability of inclusion are all true
associations. In the case study, as typical with high-throughput
genomic data, where there is a high degree of multicollinearity among
the covariates, different MCMC runs might pick different subsets
of the predictors, as variables that are highly correlated act as
proxies for each other and would be picked by different chains. This
behavior is, in general, independent of the chosen specification of the
$\alpha$ parameter.

In the case study we have applied a heavy filtering of the CGH probes.
Filtering and/or dimension reduction methods are often used in
applications of HMM models to CGH data; see, for instance, \citet
{Du:2010,Fox:2009,Guha:2008,Costa:2013}. Caution is necessary when
applying such preprocessing steps, as they may result in large gaps
between probes, thus decreasing the dependence between adjacent probes
and/or inducing heterogeneity in the gap size. In order to assess
whether the HMM approach is indeed beneficial, we looked at results on
the estimation of $\xi$ without the HMM formulation. For this we
considered the counts across the four different states as arising from
a multinomial distribution and assumed a Dirichlet hyperprior. As we
did with the HMM setting, we set all hyperparameters of the Dirichlet
to $1$. We obtained state
specific means $[ -0.25, -0.03, 0.14, 3.54 ]$ and variance estimates
$[0.41, 0.19, 0.41, 0.78]$. The HMM formulation instead resulted in
estimated means that were closer to the theoretical values as well as
in lower variance estimates (results reported on page~25). In addition,
looking at the distributions of the estimated states, the HMM approach
resulted in a larger number of neutral states, whereas the no-HMM model
classified many of these as
single copy number gains. Given the biological evidence that neutral
states should be more common, we believe this suggests that the
performance of the HMM formulation is superior despite the heavy
filtering applied to the data. A possible improvement of our HMM model
could be to incorporate the distances between adjacent probes in the
evaluation of the transition matrix, to account for possible
heterogeneity in the gap size, as done in \citet{Colella:2007} and
\citet
{Wang:2007}.

Other improvements of our model include the use of indicator variables
to model the CNV effects, in order to relax the assumption of a linear
association of the $\bolds{\xi}$ categories on the $Y$s. This would lead
to a 2-fold increase (with four categories) in the dimension of the
matrix of predictors, therefore increasing computational times.
Finally, although we have focused on array CGH data, the proposed
method can easily be extended to CNV detection using genome-wide SNP
arrays. This can be done by modifying the emission distributions in the
HMM and modeling the log-intensity ratios in equation (\ref{modelCGH})
as a mixture of uniform and normal distributions, as in \citet
{Wang:2007} and \citet{Colella:2007}.

\begin{supplement}[id=suppA]
\stitle{Supplement to ``A hierarchical Bayesian model for
inference of copy number variants and their association to gene expression''\\}
\slink[doi]{10.1214/13-AOAS705SUPP} 
\sdatatype{.pdf}
\sfilename{aoas705\_supp.pdf}
\sdescription{Description of the MCMC steps and additional results on
the case study.}
\end{supplement}

%
%

%


\printaddresses

\end{document}